\documentclass{article}
\input{tcilatex}
\begin{document}

\title{Comment on \textquotedblleft Relativity of quantum states in
entanglement swapping: Violation of Bell's inequality with no
entanglement\textquotedblright\ (arXiv:1806.02407v2 [quant-ph]; Physics
Letters A, Volume 384, Issue 15, 29 May 2020)}
\author{Luiz Carlos Ryff \\
%EndAName
\textit{Instituto de F\'{\i}sica, Universidade Federal do Rio de Janeiro,}\\
\textit{Caixa Postal 68528, 21041-972 Rio de Janeiro, Brazil}\\
E-mail: ryff@if.ufrj.br}
\maketitle

\begin{abstract}
In a recent interesting article Chris Nagele, Ebubechukwu O. Ilo-Okeke,
Peter P. Rohde, Jonathan P. Dowling, and Tim Byrnes discuss an entanglement
swapping experiment using "a setup where it is possible to switch the time
ordering of measurements." I would like to draw your attention to the fact
that the very same idea was introduced in two previous papers, and briefly
address some important points related to the subject

\end{abstract}

In a recent paper Chris Nagele, Ebubechukwu O. Ilo-Okeke, Peter P. Rohde,
Jonathan P. Dowling, and Tim Byrnes discuss the \textquotedblleft Relativity
of quantum states in entanglement swapping\textquotedblright\ \textrm{[1]}.\
According to the authors, the difference of their scenario to past works
\textquotedblleft is that the order of the measurements depends upon the
reference frame, due to relativity of simultaneity.\textquotedblright\ They
also emphasize that \textquotedblleft Other works have investigated special
relativistic effects on entanglement swapping and on teleportation, but none
have considered a setup where it is possible to switch the time ordering of
measurements.\textquotedblright\ In fact, the very same subject was
addressed in two previous papers, entitled \textquotedblleft Einstein,
Podolsky, and Rosen correlations, quantum teleportation, entanglement
swapping, and special relativity\textquotedblright\ \textrm{[2]}, and
\textquotedblleft Quantum teleportation and entanglement swapping viewed
from different moving frames\textquotedblright\ \textrm{[3]}, respectively.
In these two papers we have an experiment which can be seen as being on
ordinary Einstein-Podolsky-Rosen correlations, on quantum teleportation, or
on entanglement swapping, depending on the Lorentz frame we use to observe
it. Further, according to Nagele and coauthors, \textquotedblleft In a
moving frame, the order of the measurements is reversed, and a Bell
violation is observed even though the particles are not entangled, directly
or indirectly, or at any point in time\textquotedblright . Actually, the
situation they consider is analogous to the one considered in another paper 
\textrm{[4]} following a suggestion by Peres \textrm{[5]}. In fact, as has
been shown in ref. \textrm{[3]} and \textrm{[6]}, there is a simple
explanation for this apparently mysterious fact. The point is that whenever
the polarizations are measured first (\textquotedblleft polarization
analysis\textquotedblright ), the two remaining photons are projected into
well-defined polarization states. Consequently, when they arrive at the beam
splitter we will perform an experiment (\textquotedblleft Bell-state
analysis\textquotedblright )\ similar to the one performed by Hong, Ou, and
Mandel \textrm{[7]}, but in which the photons can have different
polarizations. Since the probability of coincident/non-coincident detection
depends on the relative polarization of the photons, by selecting the events
in which we have coincidence/no-coincidence, we obtain a subset that behaves
as if it consisted of entangled pairs of distant particles. The point is
that the probability of coincident/no-coincident detection will depend on
the experimental outcome at the polarizers. Therefore, nothing paradoxical
or more mysterious than quantum nonlocality itself needs to be invoked.

Experiments for which different interpretations are possible, depending on
the Lorentz frame we use to describe them, raise the question of what the
correct interpretation would be, or even whether it would make sense to ask
that question. Related to this point, it is interesting to consider
situations involving time-like events. As was emphasized, in this case there
is strong evidence that a measurement performed on one of the photons of an
entangled pair changes the state of the other \textrm{[8]}, in other words,
there seems to be an undeniable cause and effect relationship. As this
influence must still be present when space-like events are considered, since
the very same correlations can be observed, the idea entertained by Bohm and
Bell \textrm{[9]}\ that there must be a preferred frame does not sound so
preposterous. A way of trying to make this idea compatible with the Lorentz
transformations is to assume that the Lorentz symmetry is broken in the case
of quantum non-locality \textrm{[8]}, namely, there would be no equivalence
between active and passive Lorentz transformations.


\begin{thebibliography}{9}
\bibitem{1} Chris Nagele, Ebubechukwu O. Ilo-Okeke, Peter P. Rohde, Jonathan
P. Dowling, Tim Byrnes, \textquotedblleft Relativity of quantum states in
entanglement swapping,\textquotedblright\ Physics Letters A, Volume \textbf{%
384}, Issue 15, (29 May 2020); arXiv:1806.02407v2 [quant-ph].

\bibitem{2} Luiz C. Ryff, \textquotedblleft Einstein, Podolsky, and Rosen
correlations, quantum teleportation, entanglement swapping, and special
relativity,\textquotedblright\ Phys. Rev. A, \textbf{60}, 5083 (1999).

\bibitem{3} Luiz C. Ryff, \textquotedblleft Quantum teleportation and
entanglement swapping viewed from different moving
frames,\textquotedblright\ JOURNAL OF MODERN OPTICS, VOL. \textbf{48}, NO.
5, 905-913 (2001).

\bibitem{4} T. Jennewein, G. Weihs, J-W. Pan, and A. Zeilinger,
\textquotedblleft Experimental nonlocality proof of quantum teleportation
and entanglement,\textquotedblright\ Phys. Rev. Lett. 88, 017903 (2002).

\bibitem{5} A. Peres, \textquotedblleft Delayed choice for entanglement
swapping,\textquotedblright\ Journal of Modern Optics 47, 139 (2000).

\bibitem{6} Luiz C. Ryff, Comment on \textquotedblleft Experimental
Nonlocality Proof of Quantum Teleportation and Entanglement
Swapping,\textquotedblright\ arXiv: 0303082v1 [quant-ph].

\bibitem{7} C. K. Hong, Z. Y. Ou, and L. Mandel, \textquotedblleft
Measurement of subpicosecond time intervals between two photons by
interference,\textquotedblright\ Phys. Rev. Lett. 59, 2044 (1987).

\bibitem{8} Luiz C. Ryff, \textquotedblleft Einstein-Podolsky-Rosen (EPR)
Correlations and Superluminal Interactions," arXiv: 1506.07383v6 [quant-ph].

\bibitem{9} Interviews with J. S. Bell and D. Bohm in Davies, P.C.W., Brown,
J.R. (eds.): \textquotedblleft The Ghost in the Atom,\textquotedblright\
Cambridge University Press, Cambridge (1989).
\end{thebibliography}
\end{document}